\documentstyle[12pt,epsfig]{article}
\begin{document}
\setlength{\baselineskip}{0.30in}
\newcommand{\be}{\begin{eqnarray}}
\newcommand{\ee}{\end{eqnarray}}
\newcommand{\bi}{\bibitem}
\newcommand{\nue}{\nu_e}
\newcommand{\num}{\nu_\mu}
\newcommand{\nut}{\nu_\tau}
\newcommand{\nus}{\nu_s}
\newcommand{\mne}{m_{\nu_e}}
\newcommand{\mnm}{m_{\nu_\mu}}
\newcommand{\mnt}{m_{\nu_\tau}}
\newcommand{\munu}{\mu_{\nu}}
\newcommand{\lar}{\leftarrow}
\newcommand{\rar}{\rightarrow}
\newcommand{\lrar}{\leftrightarrow}
\newcommand{\nuh}{\nu_h}
\newcommand{\mnut}{m_{\nu_\tau}}
\newcommand{\mnh}{m_{\nu_h}}
\newcommand{\taut}{\tau_{\nut}}
\newcommand{\fg}{f_{\gamma}}
\newcommand{\mnu}{m_{\nu}}
\newcommand{\taunu}{\tau_{\nu}}
\newcommand{\nc}{\newcommand}
\newcommand{\dm}{\delta m^2}

\newcommand{\so}{\, \mbox{sin}\Omega}
\newcommand{\co}{\, \mbox{cos}\Omega}
\newcommand{\mnus}{m_{\nu_s}}
\newcommand{\taus}{\tau_{\nu_s}}
\newcommand{\cost}{\cos \theta}
\newcommand{\sint}{\sin \theta}
\newcommand{\stw}{\sin 2 \theta}
\newcommand{\ctw}{\cos 2 \theta}
\newcommand{\sv}{\sin 2 \theta}
\newcommand{\nua}{\nu_a}

\newcommand{\om}{\omega}
\newcommand{\ds}{\partial \!  \! \! /}
\newcommand{\Zs}{Z \! \! \! \! /}
\newcommand{\Ws}{W \! \! \! \! \! /}

\newcommand{\raa}{\rho_{aa}}
\newcommand{\rss}{\rho_{ss}}
\newcommand{\rsa}{\rho_{sa}}
\newcommand{\ras}{\rho_{as}}

%

%{\hbox to\hsize{April, 2002  \hfill INFN-2002}
\begin{center}
\vglue .06in
{\Large \bf { 
Cosmological Implications of Neutrinos
  }
}
\bigskip
\\{\bf A.D. Dolgov}
 \\[.05in]
{\it{INFN, sezzione di Ferrara, via Paradiso, 12, 44100 - Ferrara,
Italy \\
and\\
%\footnote
{ITEP, Bol. Cheremushkinskaya 25, Moscow 113259, Russia.}
}}\\
\end{center}
\begin{abstract}
The lectures describe several cosmological effects produced by
neutrinos. Upper and lower cosmological limits on neutrino mass are
derived. The role that neutrinos may play in formation of large scale
structure of the universe is described and neutrino mass limits are
presented. Effects of neutrinos on cosmological background radiation
and on big bang nucleosynthesis are discussed. Limits on the number of 
neutrino flavors and mass/mixing are given.
\end{abstract}

\section{Introduction \label{s-intr} }

Of all known particles neutrinos have the weakest interactions and
the smallest possibly nonvanishing, mass. Thanks to these
properties neutrino is the second most abundant particle in the
universe after photons. According to
observations the number density of photons in cosmic microwave
background radiation (CMBR) is $n_\gamma = 412$/cm$^3$. 
In standard cosmology the number density of cosmic neutrinos
can be expressed through $n_\gamma$ as
\be
n_\nu + n_{\bar \nu} = 3 n_\gamma /11 = 112/{\rm cm}^3
\label{nnu}
\ee
for any neutrino flavor ($\nue$, $\num$, and $\nut$),
assuming that there is an equal number of neutrinos and
antineutrinos. 

Knowing the temperature of CMBR, $T_\gamma = 2.728\, {\rm K} = 2.35
\cdot 10^{-4}$ eV, one can calculate the temperature of cosmic
neutrinos:
\be
T_\nu = (4/11)^{1/3} T_\gamma = 1.95\,{\rm K} = 1.68\cdot 10^{-4} {\rm
eV}
\label{tnu}
\ee
which is true if the neutrino mass is much smaller than their
temperature, $m_\nu \ll T_\nu$. Otherwise the parameter $T_\nu$ does
not have the meaning of temperature; up to a constant factor it can be 
understood as the inverse cosmological scale factor $a(t)$. 
Theory predicts that the spectrum of cosmic neutrinos, even massive 
ones, is given by the almost equilibrium form:
\be
f_\nu = \left[\, \exp\left(p/T - \xi\right) + 1 \right]^{-1}
\label{fnu}
\ee
with the dimensionless chemical potential $\xi = \mu /T$ usually
assumed to be negligibly small. 
However one should note that in the expression above $p$ is the 
neutrino momentum, while in the equilibrium distribution there stands
energy $E= \sqrt{m^2_\nu + p^2}$. 

There is a small correction to expression (\ref{fnu}) 
of the order of $(m_\nu /T_d)^2$ where $T_d$ is the
neutrino decoupling temperature, $T_d \sim {\rm MeV}$ - this is the
temperature when neutrinos stopped to interact with primeval plasma.
This correction appeared because 
at $T>T_d$ neutrinos were in equilibrium and their
distribution depended on $E/T$. Distribution of noninteracting
neutrinos should be a function of $p a(t)$. In the case of
instantaneous decoupling it turns into 
$f(\sqrt{(p/T)^2 + (m_\nu/T_d)^2})$, while for non-instantaneous
decoupling the dependence on mass could be different. 

More details about cosmological neutrinos can be found 
e.g. in a recent review paper~\cite{dolgov02}.

Neutrinos are normally assumed to possess only usual weak interactions 
with $(V-A)$-coupling to $W$ and $Z$ bosons. Correspondingly, if
$\mnu=0$, only left-handed neutrinos, $\nu_L$, i.e. those with spin 
anti-parallel to their momentum 
(and parallel for $\bar\nu$), possess this 
interaction, while right-handed neutrinos, $\nu_R$, are sterile. If
$\mnu \neq 0$ right-handed neutrinos would be also coupled to
intermediate bosons with the strength suppressed as $(\mnu/E)^2$ and
their cosmological number density would be always negligible since their
mass is bounded from above by a few eV see eqs. (\ref{gz},\ref{gz-2}).
If however neutrinos are mixed and massive (possibly with Majorana and
Dirac masses) additional three sterile neutrinos could be abundantly
produced in the early universe~\cite{dolgov81}.

It is known from experiment
that there are at least  three neutrino families (or flavors), $\nue$, 
$\num$, and $\nut$. From LEP data the number of light neutrino flavors
with $m_\nu < m_Z/2$ is indeed three:
\be
N_\nu = 2.993 \pm 0.011
\label{nnu-lep}
\ee  
One can find references to original experimental papers in the Review
of Particle Physics~\cite{pdg}. 

Direct experiment limits on neutrino masses are~\cite{pdg}
\be
\mne < 3\, {\rm eV},\,\,\,
\mnm < 190\, {\rm keV},\,\,\,  
\mnt < 18.2\, {\rm MeV} 
\label{nu-masses}
\ee
As we will see below, cosmology allows to derive an upper limit on
masses of all neutrino flavors similar to that presented above 
for $\mne$.

There is a strong evidence in favor of neutrino oscillations. The best 
fit solutions to the observed neutrino anomalies indicates maximum mixing
between $\num$ and $\nut$ with mass difference about 
$3\times 10^{-3}$ eV$^2$ (for explanation of atmospheric anomaly)
and also large mixing between $\nue$ and another active neutrino with
mass difference between $10^{-3}-10^{-5}$ eV$^2$ (for explanation of
the deficit of solar neutrinos). 
If mass differences are indeed so small then masses of all active
neutrinos should be below 3 eV and right-handed neutrinos would not be
practically produced in particle interactions in the standard theory,
but as we noted below, they may be produced by oscillations. 

Except for the above mentioned anomalies,
and possibly LSND, neutrinos are well described by the standard
electroweak theory. For a recent review of neutrino anomalies see
e.g. ref.~\cite{nu-osc}.

In what follows we discuss the bounds in neutrino masses that can be
derived from the magnitude of cosmic energy density and large scale 
structure of the universe (sec. \ref{s-mass}).
Relation between cosmological neutrinos and CMBR is considered in
sec. \ref{s-cmbr}.
In section \ref{s-bbn}
we describe the role played by neutrinos in big bang
nucleosynthesis (BBN) and present the limits on the number of neutrino
species and possible neutrino degeneracy.
Cosmological impact of neutrino oscillations is considered in 
sec. \ref{s-nuosc}.
The body of the lectures is preceded by a brief presentation of basic
cosmological facts and essential observational data 
(section~\ref{s-cosmo}). These lectures present a shorter version of
the recent review paper~\cite{dolgov02} where one can find details and
a long list or references, many of which are omitted here because of
lack of space and time.

\section{A little about cosmology \label{s-cosmo}}

The universe is known to expand according to the Hubble law:
\be
V = H r
\label{hubble}
\ee
where $V$ is the velocity of a distant object, $r$ is the distance to
it and $H=\dot a /a$ is the Hubble constant (or better to say, Hubble 
parameter,
since it is not constant in time). The present day value of $H$ is
given by
\be
H = 100\, h\,{\rm km/sec/Mpc}
\label{H}\
\ee
with $h = 0.7 \pm 0.1$. There are still indications for smaller and
larger values of $H$ but we will not go into details here. One can
find discussion of determination and values of this and other
cosmological parameters e.g. in recent papers~\cite{prmtr}.

The critical or closure energy density is proportional to $H^2$ and is
equal to:
\be
\rho_c = {3H^2 m_{Pl}^2 \over 8 \pi} = 10.6\,h^2 \, {\rm keV /cm^3}
\label{rhoc}
\ee
where $m_{Pl}= 1.221$ GeV is the Planck mass.
Contributions of different forms of matter into cosmological energy
density is usually presented in terms of dimensionless parameter 
$\Omega_j = \rho_j /\rho_c$. According to the data the dominant part
of cosmological energy density is given either by vacuum energy or by
an unknown form of matter which has negative pressure, 
$p <-\, \rho/3$, and induces 
an accelerated expansion (anti-gravity) at the present epoch.
Its energy density is $\Omega_{vac} \approx 0.7$. The total energy
density is close to the critical value so $\Omega_{tot} \approx 1$. The
contribution of the usual baryonic matter, as
determined from CMBR, is roughly $\Omega_b h^2 = 0.022$. This is
is consistent with determination of $\Omega_b$ from BBN. The remaining
0.25 is believed to be contributed by some unknown elementary particles
(though, say, black holes are not excluded) weakly interacting with 
photons - that's why they are called dark or invisible matter. 

We are interested in a rather late period of the universe evolution 
when the temperature was in MeV range or below down to the present 
time. For more details about cosmology one can see any textbook or 
e.g. the recent reviews~\cite{dolgov02,pdg}. Initially all the 
particles in primary plasma, photons,
$e^+e^-$-pairs, three flavors of neutrinos and antineutrinos, and a
little baryons were in strong thermal contact and hence had equilibrium
distributions (\ref{fnu}) for fermions and similar expressions with 
minus sign in front of 1 for bosons. Particle energy $E$ should stand in
this equation instead of $p$ but since majority of particles are
relativistic, this difference is not important. The energy density of 
massless particles in thermal equilibrium is given by the expression:
\be
\rho = {\pi^2\,g_* T^4/ 30}
\label{rhoin}
\ee
where $g_* = 10.75$ includes contribution from all mentioned above
particles except for neglected baryons. 

At that stage the energy density was almost precisely equal to the
critical one, with accuracy better than $10^{-15}$,
the particles were relativistic (for $T>m_e$) with
equation of state $p=\rho/3$, where $p$ is the pressure density, and
thus:
\be
H = {1\over 2t} = \left({ 8\pi^3 g_* \over 90}\right)^{1/2}\,
{T^2 \over m_{Pl}} = 5.443 \left({g_* \over 10.75}\right)^{1/2}
{T^2 \over m_{Pl}}
\label{HofT}
\ee

Cross-section of neutrino interactions behaves as 
$\sigma_\nu \sim G_F^2 E^2$ and the reaction rate is
$\dot n/n \sim \sigma_\nu n \sim G_F^2 T^5$. Here $n \sim T^3$
is the particle number density and $G_F = 1.166\cdot 10^{-5}$ GeV$^2$
is the Fermi coupling constant. Comparing reaction rate with the
Hubble expansion rate we can conclude that at temperatures above a few 
MeV neutrinos should be in thermal equilibrium.
To be more precise one should consider kinetic equation
governing neutrino distribution in cosmological background:
\be
\left( \partial_t + H p \partial_p \right) f_\nu (t,p)= 
Hx \partial_x f_\nu (x,y)= I_{coll}
\label{kineq}
\ee 
where $p$ is the neutrino momentum, $x = 1/a(t)$, $y = pa /m_0$,
$a(t)$ is the cosmological scale factor, and $m_0$ is the
normalization mass which we take as $m_0 =1$ MeV. On relativistic
stage when $T \sim 1/a$ is convenient to take $x= m_0 /T$ and
$y= p/T$. 

For an estimate of neutrino decoupling temperature we neglect inverse
reactions in the collision integral and assume Boltzmann statistics.
Each of the neglected effects would enlarge the decoupling temperature
by about 10-15\%. In this approximation the kinetic equation becomes
\be
Hx\, {\partial f_\nu \over f_\nu \partial x } = 
- {D\,G_F^2 y \over 3 \pi^3 x^5}
\label{hxdf}
\ee
where $D$ is a constant. Usually in the estimates of decoupling
temperature one takes thermally average value of neutrino momentum,
$\langle y \rangle = 3$. If we include all possible reactions where
neutrinos may participate then $D = 80 (1+g_L^2+g_R^2)$ with
$g_L = \sin^2 \theta_W \pm 1/2$ and $g_R = \sin^2 \theta_W$ where
$\sin^2 \theta_W =0.23$ and
the sign $''+''$ stands for $\nue$ and $''-''$ stands for
$\nu_{\mu,\tau}$. Correspondingly the decoupling temperature
determined with respect to the total reaction rate would be
$T^d_{\nue}=1.34$ MeV and $T^d_{\num,\nut} = 1.5 $ MeV for $\nue$ and
$\nu_{\mu,\tau}$ respectively.

If we take into account only $(\nu-e)$-interactions then 
$D= 80 (g_L^2+g_R^2)$ and the decoupling
of neutrinos from electron-positron (and photon) plasma would take
place at $T^d_{\nue}=1.87$ MeV and $T^d_{\num,\nut} = 3.12 $ MeV. 
Above $T^d$ the temperatures of photon-electron-positron and neutrino 
plasma should be equal. In fact in the standard model equality is 
maintained down to $T\approx m_e$ (see below).

If only $\nu \bar \nu$-annihilation into $e^+e^-$ is taken into
account then $D= 16 (g_L^2+g_R^2)$ and the decoupling temperatures
would be $T^d_{\nue}=3.2$ MeV and $T^d_{\num,\nut} = 5.34 $ MeV.
Below these temperatures the total number density of neutrinos in
comoving volume could not change.

When temperature dropped below electron mass, $e^+e^-$-pairs
annihilated heating photons but leaving neutrinos intact. As a result
of this heating $T_\gamma$ become higher than $T_\nu$,
eq.~(\ref{tnu}), and relative neutrino number density dropped with 
respect to photons as
given by eq.~(\ref{nnu}) instead of earlier existed equilibrium ratio
$(n_\nu + n_{\bar \nu})/n_\gamma = 3/4$.

\section{Cosmological limits on neutrino mass
\protect\footnote{More detailed discussion of such limits can
be found in the recent reviews~\cite{dolgov02,kainulainen02}.}\protect
\label{s-mass}}

\subsection{Gerstein-Zeldovich limit \label{ss-gz}}

Since the number density of neutrinos at the present day is known,
see eq.~(\ref{nnu}), it
is easy to calculate their contribution into cosmological energy
density, $\rho_\nu = \sum m_{\nu_a} n_\nu$, if neutrinos are 
stable. Demanding that $\rho_\nu$
does not exceed the known value of energy density of matter we obtain
\be
\sum_a m_{\nu_a} < 95\,{\rm eV}\, \Omega_m h^2 \approx 14\,{\rm eV}
\label{gz}
\ee
where the sum is taken over all light neutrino species, 
$ a=e,\,\mu,\,\tau$. 
This limit was originally derived by Gerstein and Zeldovich~\cite{gz}
in 1966. Six years later the result was rediscovered by Cowsik and
McClelland~\cite{cm}. In the later paper, however, the photon heating
by $e^+e^-$-annihilation was not taken into account and both helicity
states of massive neutrinos were assumed to be equally abundant. 
Correspondingly the resulting number density of relic neutrinos was 
overestimated by the factor 11/2. 

If all active neutrinos are strongly mixed and their mass differences
are very small (see the end of sec.~\ref{s-intr}) then the limit
(\ref{gz}) for an individual mass would be $\mnu < 4.7$ eV.

The bound (\ref{gz}) can be noticeably strengthened because neutrino 
may make only sub-dominant contribution to $\Omega_m$. Arguments
based on large scale structure formation (see below sec.~\ref{ss-lss}) 
lead to the conclusion that $\Omega_\nu < \Omega_m/3$ and
correspondingly: 
\be
\sum_a m_{\nu_a} < 5 \,\,{\rm eV}
\label{gz-2}
\ee
As noted above, in the case of small mass differences the mass bound
for a single neutrino would be $m_\nu < 1.7 $ eV.

\subsection{Tremaine-Gunn limit \label{ss-tg}}

Quantum mechanics allows to obtain a lower limit on neutrino mass if
neutrinos make {\it all} dark matter in galaxies, especially in dwarf
ones~\cite{tremaine79}. The derivation is based on the fact that
neutrinos are fermions and hence cannot have an arbitrary large number
density if their energy is bounded from above to 
allow formation of gravitationally
bound cluster. So to make dominant contribution into dark matter
neutrino mass should be larger than a certain value. Gravitationally 
bound neutrinos would be most
densely packed if they form degenerate gas with Fermi momentum
$p_f = m_\nu V_F$. The Fermi velocity $V_F$ can be determined
from the virial theorem:
\be
V_F^2 = G_N M_{gal} /R_{gal}
\label{vf}
\ee
where $G_N = 1/m_{Pl}^2$ is the Newton gravitational constant and
$M_{gal}$ and $R_{gal}$ are respectively the mass and radius of a
galaxy. 

The number density of degenerate neutrinos and equal number of 
antineutrinos is $n_\nu = p^3_F/(3\pi^2)$
and correspondingly their total mass in a galaxy is
\be
M_\nu = 4\pi R_{gal}^3 m_\nu n_\nu /3
\label{Mnu}
\ee
According to observations galactic masses are dominated by invisible 
matter, so one should expect that
$M_\nu \approx M_{gal}$. From the equations above we find:
\be
m_\nu = 80\,{\rm eV}\,\left({300\,{\rm km/sec} \over V }\right)^{1/4}
\left({ 1 \, {\rm kpc} \over R_{gal}}\right)^{1/2}
\label{mnutg}
\ee
For dwarfs $R_{gal} \approx 1$ kpc and $V \approx 100$
km/sec. Correspondingly neutrinos, if they constitute all dark matter
in such galaxies, should be rather heavy, $m_\nu > 100$ eV in
contradiction with Gerstein-Zeldovich limit. Thus we have to conclude
that dark matter in galaxies is dominated by some other unknown 
particles.

\subsection{Neutrinos and large scale structure of the universe
\label{ss-lss}} 

Though, as we saw above, massive neutrinos cannot be dominant dark 
matter particles, they may play an essential role in large scale
structure formation and evolution. According to the accepted point of
view cosmological structures have been developed as a result of
gravitational instability of initially small primordial density
perturbations. The latter presumably were generated at inflationary
stage due to rising quantum fluctuations of the inflaton field. 
For reviews and list of references see e.g.~\cite{bellido00}.
It is usually assumed that the spectrum of initial density
perturbation has a simple power law form, i.e. Fourier transform
of the density perturbations 
\be
(\delta \rho /\rho)_{in} =\int d^3 k \,\delta (k)
\label{deltarho}
\ee
behaves as  $\delta^2 \sim k^n$. Moreover, the value of the
exponent, $n$, is usually taken to be 1. It corresponds to  
flat or Harrison-Zeldovich spectrum~\cite{hz}, as indicated by 
inflation and consistent with observations.

With the known initial perturbations and equation of state of
cosmological matter one can calculate the shape of the evolved
spectrum and to compare it with observations. This permits to
determine the properties of the cosmological dark matter. In the case
of neutrinos density perturbations at small scales are efficiently
erased as can be seen from the following simple arguments. Neutrinos
were decoupled from plasma when they relativistic. The decoupling
temperature is  $T^d \sim $ MeV,
while $\mnu \leq 10$ eV. Thus after decoupling neutrinos free streamed
practically with the speed of light. Since the flux of
neutrinos from neutrino-rich regions should be larger than that from
neutrino-poor regions, the inhomogeneities in neutrino distribution
would smoothed down at the scales smaller than neutrino free path, 
$l_{fs}= 2t_{nr}$. Here $t_{nr}$ is the cosmic time from beginning
till the moment when neutrinos became nonrelativistic. As we mentioned
above neutrinos
propagate with the speed of light, so locally their path is equal just
to $t$ and factor 2 came from the expansion of the universe. The mass
contained inside $l_{fs}$ is
\be 
M_{fs} = {4 \pi (2t_{fs})^3 \over 3} \, \rho = m_{Pl}^2 t_{fs}
\label{mfs}
\ee
where we used for the cosmological energy density the critical value
(\ref{rhoc}) with the Hubble parameter $H=1/(2t)$, as given by
eq.~(\ref{HofT}). Assuming that the universe was dominated by
relativistic matter (photons and three neutrino flavors) till neutrino
temperature dropped down to $T_\nu = \mnu /3$ and taking into account
that $T_\nu \approx 0.7 T_\gamma$ (\ref{tnu}) we find that the mass
inside the free-streaming length is
\be
M_{fs} = 0.1 m_{Pl}^3 /\mnu^2 \approx
10^{17} M_\odot ({\rm eV} /\mnu)^2
\label{mfsnum}
\ee
where $M_\odot = 2\cdot 10^{33}$ g is the solar mass. This result is
derived for the case of one neutrino much heavier than the others. It
would be modified in an evident way if neutrinos are mass degenerate.

In such a theory the characteristic mass of the first formed objects,
$M_{fs}$, is much larger than the mass of large galaxies, 
$M_{gal} \sim 10^{12} M_\odot$ and dark matter with such property is
called hot dark matter (HDM). The dark matter particles for which
the characteristic mass is smaller than the galactic
mass are called cold dark matter (CDM) and the intermediate case is
naturally called warm (WDM). In HDM model of structure formation large
clusters of galaxies should be formed first and smaller structures
could be created from larger ones later by their fragmentation. 
However such process demands too much time and, moreover, the 
observations indicate that smaller structures are older. Together with
Tremaine-Gunn limit discussed in sec.~\ref{ss-tg}, it ``twice''
excludes neutrinos as dominant part of dark matter in the universe.
However mixed models with comparable amount of CDM and HDM are not
excluded. Though the mystery of cosmic conspiracy - why different
particles have comparable contribution to $\Omega$ - becomes in this
case even more pronounced: 
\be
\Omega_{vac}\sim (\Omega_m =\Omega_{CDM}+\Omega_{HDM})
\sim \Omega_b
\label{Omega-a}
\ee

From the arguments presented here one can see that the larger is the
fraction of neutrinos in the total mass density of the universe the
smaller should be power in cosmic structures at small scales. This 
permits to strengthen the upper limit on neutrino mass. 
Especially sensitive to neutrino mass are the structures at large
red-shift $z$ because in neutrino dominated universe small structures 
should form late and should not exist at large $z$. The neutrino
impact on the structure formation was analyzed in refs.~\cite{nustr}
with the typical limits between 1 and 5 eV. More detailed discussion 
and more references can be found in the review~\cite{dolgov02}.
According to ref.~\cite{sdss99} Sloan Digital Sky Survey is
potentially sensitive to $\mnu \leq 0.1$ eV. 

\subsection{ Cosmological limit on heavy neutrino mass \label{ss-nuh}}

If there exists fourth lepton generation then the corresponding neutrino
should be heavier than $m_Z/2 = 45$ GeV to surpass the LEP
result~(\ref{nnu-lep}). If these heavy neutrinos are stable on
cosmological time scale, $\tau_\nu \geq t_U \sim 10^{10}$ years, then
their mass density may be cosmologically noticeable. Since such
neutrinos are assumed to be very heavy their number density at
decoupling should be Boltzmann suppressed and they may escape
Gerstein-Zeldovich limit. First calculations of cosmological number
density of massive particles were performed by Zeldovich in
1965~\cite{zeldovich65}. However his result contained a numerical
error later corrected in ref.~\cite{zop65}. The same approach was
applied to the calculations of the number/energy density of relic
heavy neutrinos practically simultaneously in the papers~\cite{vdz}
where it was found that the mass of heavy neutrino should be above 2.5
GeV to be cosmologically safe. 

The number density of massive particle (neutrinos)s which survived 
annihilation is inversely proportional to the annihilation 
cross-section $\sigma_{ann}$ and is approximately given by the 
expression
\be
n_\nu / n_\gamma  = \left( \sigma_{ann}\,v\,m_\nu m_{Pl} \right)^{-1}
\label{nmnu}
\ee
where $v$ is the c.m. velocity of the annihilating particles and
$n_\gamma$ is the number density of photons in CMBR. For relatively
light neutrinos, $\mnu \ll m_Z$ (which is not realistic now), the
annihilation cross-section is proportional to $\sigma \sim \mnu^2$ and
the energy density of heavy relic neutrinos drops as
$1/\mnu^2$. According to the calculations quoted above 
$\rho_\nu = \rho_c$ for $\mnu = 2.5 $ GeV. 

For higher masses $\mnu > m_{W,Z}$, the cross-section started to drop
as $\sigma \sim \alpha^2 / \mnu^2$ and the cosmologically allowed 
window above 2.5 GeV becomes closed for $\mnu >$ (3-5) TeV~\cite{dz80}. 
However for $\mnu >m_W$ a new channel of annihilation becomes open,
\be
\nuh+\bar\nuh \rar W^+W^-
\label{nuW}
\ee
with the cross-section rising as
$m_{\nuh}$~\cite{ekm89}. The rise of the cross-section is related to
the rise of the Yukawa couplings of Higgs boson which is necessary to
ensure a large mass of $\nuh$. Correspondingly the excluded region
above a few TeV becomes open again. However the annihilation (\ref{nuW})
proceeds only in one lowest partial wave and the cross-section is
restricted by the unitarity limit~\cite{griest90},
\be
\sigma_J < \pi (2J+1)/p^2.
\label{sigma0}
\ee
If one assumes that this limit is
saturated then the large values $m_{\nuh}$ about 100 TeV would be
forbidden. In reality the limit should be somewhat more restrictive
because it is natural to expect that the cross-section started to drop
with rising mass of neutrino before it reaches the unitarity
bound. However it is very difficult, if possible at all, to make
any accurate calculations in this strong interaction regime.

\begin{figure}[htb]
\begin{center}
  \leavevmode
  \hbox{
    \epsfysize=3.0in
    \epsffile{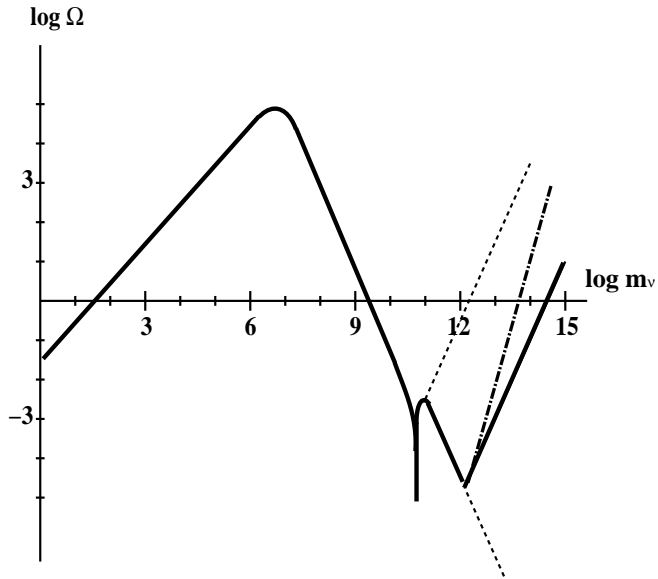}}
\end{center}
\caption{Cosmological energy density of massive neutrinos
$\Omega = \rho_{\nuh} /\rho_c$ as a function of their mass 
measured in eV. The meaning of different lines is explained
in the text.
\label{rholfig}}
\end{figure}

To summarize this discussion, the cosmic energy density, $\rho_{\nuh}$,
of heavy neutrinos with the usual weak interaction
is sketched in fig. (\ref{rholfig}). In the region of very small masses the
ratio of number densities $n_{\nuh}/n_\gamma$ does not depend upon the
neutrino mass and $\rho_{\nuh}$ linearly rises with mass. For larger masses
$\sigma_{ann} \sim \mnh^2$ and $\rho_{\nuh}\sim 1/\mnh^2$. This formally opens
a window for $\mnh$ above 2.5 GeV. A very deep minimum in $\rho_{\nuh}$ near
$\mnh = m_Z /2$ is related to the resonance enhanced cross-section around
$Z$-pole. Above $Z$-pole the cross-section of $\bar \nuh \nuh$-annihilation
into light fermions goes down with mass as $\alpha^2/\mnh^2$ (as in any normal
weakly coupled gauge theory). The corresponding rise in $\rho_{\nuh}$ is
shown by a dashed line. However for $\mnh > m_W$ the contribution of the   
channel $\bar \nuh \nuh \rar W^+W^-$ leads to the rise of the cross-section
with increasing neutrino mass as $\sigma_{ann} \sim \alpha^2 \mnh^2 /m_W^4$.
This would allow keeping $\rho_{\nuh}$ well below $\rho_c$
for all masses above
2.5 GeV. The behavior of $\rho_{\nuh}$, with this effect of rising
cross-section included, is shown by the solid line up to
$\mnh =1.5 $ TeV. Above that value it continues as a dashed line.
This rise with mass would break unitarity limit for partial wave
amplitude when $\mnh$ reaches 1.5 TeV (or 3 TeV for Majorana neutrino).
If one takes the maximum value of the S-wave cross-section permitted by
unitarity (\ref{sigma0}), which scales as $1/\mnh^2$, this would give rise 
to $\rho_{\nuh} \sim \mnh^2$ and it crosses $\rho_c$ at $\mnh \approx 200$ TeV.
This behavior is continued by the solid line above 1.5 TeV.
However for $\mnh \geq {\rm a\,\, few}\,\, {\rm TeV}$ the Yukawa coupling of  
$\nuh$ to the Higgs field becomes strong and no reliable calculations
of the annihilation cross-section has been done in this limit.
Presumably the cross-section is much
smaller than the perturbative result and the cosmological bound
for $\mnh$ is close to several TeV. This possible, though not certain,  
behavior is presented by the dashed-dotted line. One should keep in mind,  
however, that the presented results for the energy density could only be true
if the temperature of the universe at an early stage was higher than
the heavy lepton mass.

\section{Neutrinos and CMBR \label{s-cmbr}}

\begin{figure}[htb]
\begin{center}
  \leavevmode
  \hbox{
\epsfysize=4.0in    
\epsffile{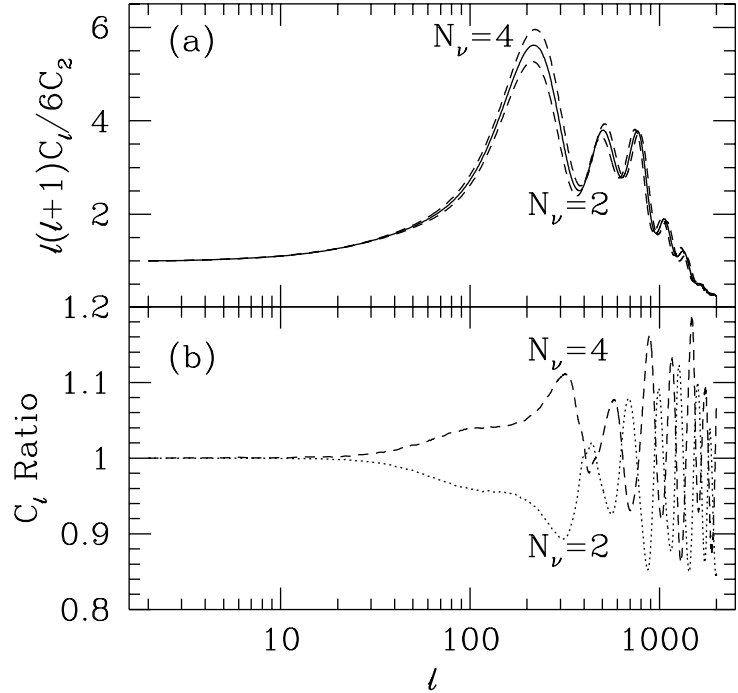}}
\end{center}
\caption{(a) An example of an angular spectrum of CMB anisotropies 
with varying
number of neutrino species, $k_\nu =2,3,4$. (b) The ratio of $C_l$ for
$k_\nu =2,\,4$ relative to $k_\nu =3$ (from ref.~\protect\cite{hssw})}
\label{figcmb}
\end{figure}

Measurements of the angular fluctuations of CMBR, which are in
continuous progress now, also permit to obtain valuable information
about cosmic neutrinos. The spectrum of fluctuations is presented in
terms of $C_l$, the squares of the amplitudes in the decomposition
of the temperature fluctuations in terms of spherical harmonics:
\be
{\Delta T \over T} = \sum_{l,m} a_{lm} Y_{lm} (\theta, \phi)
\label{spherdt}
\ee
and
\be
C_l = {1\over 2l+1} \sum_{m=-l}^l |a_{lm}|^2
\label{C-l}
\ee
A typical spectrum of fluctuations is presented in  
fig.~\ref{figcmb}(a)~\cite{hssw}. 
 
For low $l$ the amplitudes $C_l$ are practically $l$-independent if
the spectrum of initial density perturbations is flat. At 
$l\approx 200 $ it has a pronounced peak and a few weaker peaks at 
larger $l$. At $l>10^3$ the fluctuations are strongly damped. A
detailed explanation of these spectral features can be found e.g. in
the review~\cite{gawiser00}. These peaks were produced by sound waves
at the earlier stage, roughly speaking at the moment of hydrogen
recombination at $T \approx 3000$ K. After this moment the universe
became transparent to CMB photons and the features existed at the 
moment of recombination became ``frozen'' and are observed now in the
sky. The first peak of the largest amplitude corresponds to the 
``last'' sound wave with the wave length equal to the horizon size at 
recombination divided by the speed of sound which for relativistic 
plasma is $c_s =1/\sqrt{3}$:
\be 
\lambda_1 = l^{(rec)}_{hor} /\sqrt{3}
\label{lambda1}
\ee
The other peaks correspond to maximum compression or rarefaction at
the same moment and their phase is larger by $n \pi $. Their amplitude
is typically smaller because corresponding waves was generated earlier 
and had more time to decay. 

The physical scale $l^{(rec)}_{hor}$ depends upon the expansion regime
and, in particular, upon the fraction of relativistic matter. Because
of that the peak 
position is sensitive to number of neutrino families and 
to neutrino mass. However this effect is rather weak and to a much
larger extent the position of the peak is 
determined by the geometry of
the universe because the angle at which we see $l^{(rec)}_{hor}$ 
at the present time
depends upon the curvature of space. The data strongly support
spatially flat universe, $\Omega_{tot} =1$ (for an analysis of 
previous and new data see e.g. ref.~\cite{sievers02}). 

More sensitive to the contribution of relativistic matter are heights
of the peaks. The point is that at non-relativistic 
or, in other words, at matter-dominated
(MD) stage gravitational potential of perturbations, $\psi$, remains 
constant. Indeed the potential satisfies the Laplace equation: 
\be
(1/a^2)\, \Delta \psi \sim \delta \rho
\label{delta-psi}
\ee
where $a$ is the cosmological scale factor and $\delta \rho$ is the
density contrast. It is known that density perturbations at MD-stage 
rise as $\delta \rho /\rho \sim a$, while the total energy density
decreases as $\rho \sim 1/a^3$. Correspondingly
\be
\psi \sim a^2 \rho\, (\delta \rho /\rho) = const 
\label{psi}
\ee
If the cosmological expansion is not exactly non-relativistic due to
presence of some relativistic matter (neutrinos) the gravitational
potential would be time depending, $\psi = \psi (t)$, and the sound
waves (which are induced by gravity) would be amplified - the effect is
analogous to parametric resonance amplification. 

The heights of the peaks also depend upon the fraction of baryonic
matter, $\Omega_b h^2$ since the latter makes the main contribution
to the mass of acoustic oscillators, while dark matter particles do
not experience any pressure from photons of CMB. The degeneracy of the
impact of different cosmological parameters on the angular 
spectrum of CMBR makes determination of these parameters much 
more difficult and one needs to invoke additional information 
from other pieces of astronomical data
and/or to wait till more precise measurement of all $C_l$ in 
forthcoming experiments.

At the present day the accuracy of determination of the 
properties of cosmic neutrinos 
from CMBR is not very good. In ref.~\cite{hannestad01} 
the upper limit on the number of neutrino families was found
$N_\nu <17$ (95\% confidence level) for the Hubble parameter 
$h = 0.72\pm 0.08 $ and $\Omega_b h^2 = 0.020\pm 0.002$. If these 
parameters are larger a larger fraction of relativistic energy density
would be allowed and more neutrino flavors or other relativistic
particles may exist. An additional account of the data on the large 
scale structure~\cite{hannestad01} permitted to arrive to an 
interesting {\it lower} limit, $N_\nu > 1.5$. Thus an independent
indication (in addition to BBN) of non-vanishing cosmological background
of massless or very light neutrinos is obtained. Of course, these 
results are not competitive with BBN (see sec. \ref{s-bbn}) at the
present time. However they can be such in near future.

If neutrinos are massive and contribute into hot component of dark 
matter, their presence can be traced through 
CMBR~\cite{dodelson95}. Both effects mentioned above, a shift of the
peak positions and a change of their heights, manifest themselves depending
on the fraction of hot dark matter $\Omega_{HDM}$. Moreover the angular
spectrum of CMBR is sensitive also to the value of neutrino mass
because the latter shifts $t_{eq}$,
the moment of the transition from radiation dominance to matter dominance.
According to the paper~\cite{dodelson95} the amplitude of
angular fluctuations of CMBR is 5-10\% larger for $400<l<1000$
in the mixed hot-cold dark matter (HCDM) model with
$\Omega_\nu =0.2-0.3$ in comparison with the pure CDM model.
An analysis of the data of ref.~\cite{2ndpeak} on CMBR angular
spectrum was performed in the paper.~\cite{wang01} and the best-fit
range of neutrino mass was found:
\be
m_\nu = 0.04 - 2.2\,\,{\rm eV}.
\label{mnu-cmbr}
\ee

An interesting effect
which is related to neutrino physics in the early
universe when the latter was about 1 sec old and the temperature was
in MeV range may be observed in the forthcoming Planck mission if 
the expected accuracy at per cent level is achieved. It is usually
assumed that at that time neutrinos had an equilibrium spectrum with
the temperature which was initially equal to the temperature of
photons, electrons, and positrons, while somewhat later at $T< m_e$
neutrino temperature dropped with respect to the photon one 
because $e^+e^-$-annihilation heated
photons but not neutrinos since neutrinos were already decoupled from 
electrons and positrons (see eq.~(\ref{tnu}) 
and discussion at the end of sec. \ref{s-cosmo}). However
the decoupling was not an instantaneous process and some residual
interactions between $e^\pm$ and neutrinos still existed at smaller
temperatures. The annihilation of the hotter electron-positron pairs, 
$e^+ e^- \rightarrow \bar \nu \nu$, would heat up the neutrino
component of the plasma and distort the neutrino spectrum. The
average neutrino heating under assumption that their spectrum 
maintained equilibrium was estimated in ref.~\cite{dic}.
However, the approximation of the equilibrium spectrum is
significantly violated and this assumption was abolished in 
subsequent works. In the earlier papers \cite{df,dt} kinetic 
equations were approximately solved in Boltzmann 
approximation. In ref. \cite{dt} the effect was calculated 
numerically, while in ref. \cite{df} an approximate analytical
expression was derived. After correction of the numerical factor 1/2
the calculated spectral distortion has the form:
\be
{\delta f_{\nue} \over f_{\nue} }\approx 3\cdot 10^{-4} \,\,{E\over T}
\left( {11 E \over 4T } - 3\right)
\label{dff}
\ee
Here $\delta f = f - f^{(eq)}$.
The distortion of the spectra of $\num$ and $\nut$ is approximately twice
weaker. 

An exact numerical treatment of the problem (i.e. numerical solution
of the integro-differential kinetic equations without any simplifying
approximations) was conducted in the papers~\cite{hanmad}-\cite{gg}.
The accuracy of the calculations achieved in ref.~\cite{dhs0} was the
highest and some difference with the results of two other papers can
be prescribed to a smaller number of grids in the collision 
integral~\cite{hanmad} or to non-optimal distribution of
them~\cite{gg} (see discussion of different methods of
calculations in ref.~\cite{dolgov02}). 
Recently calculations of the distortion of neutrino spectrum
were done in ref.~\cite{esposito00} in a completely different way
using expansion in interpolating polynomials in 
momentum~\cite{dolgov97}. The results of this work perfectly agree
with those of ref.~\cite{dhs0}.   

One would expect that the distortion of neutrino spectrum at a per
cent level would result in a similar distortion in the primordial 
abundances of light elements. However, this does not occur because an
excess of energetic neutrinos over the equilibrium spectrum which
would give rise to a larger neutron-to-proton ratio and to a larger 
mass fraction of primordial $^4He$ is compensated by an increase of 
the total energy density of $\nue$ which acts in the opposite
direction diminishing the neutron-proton freezing temperature and thus
diminishing the $n/p$-ratio (see below sec. \ref{s-bbn}). The net
result of this distortion on $^4He$ is at the level of $10^{-4}$.
Still the observation of this small deviation of neutrinos from
equilibrium is not impossible. The corrections discussed here 
and electromagnetic corrections of ref.~\cite{ldht} could be 
interpreted as a change of  $N_\nu$ from 3 to 3.04. Planck mission may
detect this effect but the concrete features depend upon the ratio of
neutrino mass to the recombination temperature 
$T_{rec}= 3000\, {\rm K} = 0.26$ eV.

\section{Neutrinos and BBN \label{s-bbn}}

\subsection{BBN - brief description \label{ss-descr}}

Physical processes essential for BBN took place when the temperature
of the primeval plasma was in the interval from a few MeV down to
60-70 keV. According to eq.~(\ref{HofT}) the characteristic time was
respectively between 0.1 sec up to 200 sec. During this period light
elements $^2H$, $^3He$, $^4He$, and $^7Li$ were synthesized. 
As one can see from
eq.~(\ref{HofT}) the universe cooling rate i.e. time-temperature
relation depends on the number of particle species in the
primeval plasma, $g_*$.
At BBN epoch the latter is usually parameterized as
\be
g_*  = 10.75 + (7/4)(N_\nu -3)
\label{gstar}
\ee
where the effective number of additional neutrinos $\Delta N_\nu =
N_\nu -3$ describes any form of energy present during BBN. This
parameterization is precise if a non-standard energy has relativistic
equation of state, $p =\rho/3$. In any other case $\Delta N$ becomes a
function of time and, moreover, the impact of additional energy on
primordial abundances of different light elements could be different
from that created by the equivalent number of massless neutrinos.

Building blocks for for creation of light elements were prepared in
the weak interaction reactions:
\be
n+\nu_e &\lrar& p + e^-,
\label{nnue} \\
n+e^+ &\lrar& p +  \bar \nu
\label{ne}
\ee
At high temperatures, $T>0.7$ MeV, these reactions were fast in
comparison with the universe expansion rate $H$ and the neutron-proton
ratio followed the equilibrium curve:
\be
(n/p) = \exp \left( -\Delta m /T \right)\,
\exp\left( - \xi_{\nue}\right)
\label{n/p}
\ee
where $\Delta m = 1.3$ MeV is the neutron-proton mass difference and
$\xi_{\nue} = \mu_{\nue} /T$ is dimensionless chemical potential of
electronic neutrinos. At smaller $T$ reactions (\ref{nnue},\ref{ne})
became effectively frozen and the ratio $n/p$ would be constant if not
the slow neutron decay with the life-time 
$\tau_n = 885.7 \pm 0.8$~sec~\cite{pdg}. 

The temperature of the freezing of the reactions
(\ref{nnue},\ref{ne}), $T_{np}$,
is determined by the competition of the reaction rate,
$\Gamma\sim G_F^2 T^5$ and the expansion rate, 
$H \sim \sqrt{g_*}\, T^2$. Hence  
$T_{np} \sim {g_*}^{1/6}$. Larger is $g_*$, larger is $T_{np}$,
and more neutrons would remain for creation of light elements. On the
other hand, the nucleosynthesis temperature, $T_{NS}$, does not depend
upon $g_*$ but the time when $T_{NS}$ is reached depends upon it. 
Larger is $g_*$ shorter is this time. Correspondingly less neutrons
would decay and more helium-4 and deuterium would be created. Thus a
variation of $g_*$ acts in the same direction in both phenomena. In
particular, an increase of $N_\nu$ by 1 leads to an 
increase of $^4He$ by about
5\% as one can easily check using the publicly available BBN
code~\cite{kawano}. 

\begin{figure}[htb]
\begin{center}
  \leavevmode
  \hbox{
 \epsfysize=4.3in
 \epsffile{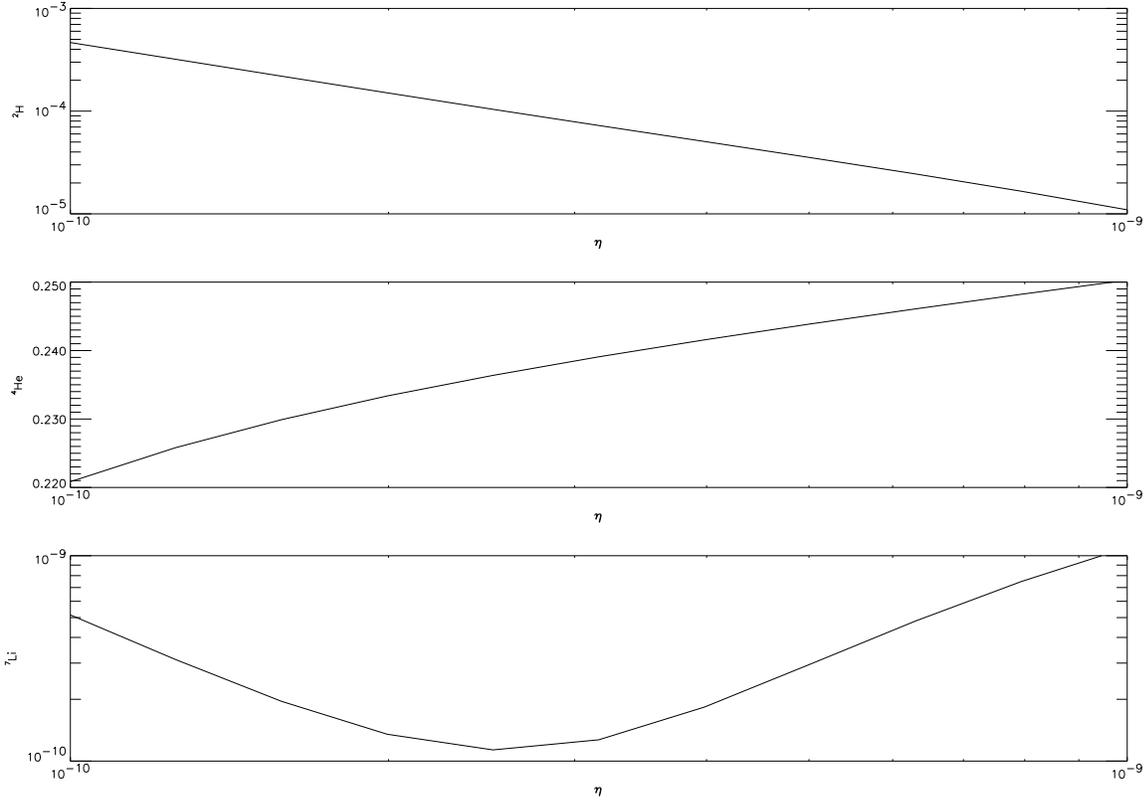}}
%\epsffile{elmntofetafig/h2ofeta.ps}}
\end{center}
\caption{Abundances of light elements $^2H$ (by number) $^4He$ (by mass),
and $^7 Li$ (by number) as functions of baryon-to-photon ratio
$\eta_{10} \equiv 10^{10}n_B/n_\gamma$.
\label{primabund}}
\end{figure}

Light elements begin to form from the primeval protons and neutrons
when the temperature reached the value
\be
T_{NS} =  { 0.064 \,{\rm MeV} \over 1 - 0.029 \ln \eta_{10}}
\label{TNS}
\ee
where $\eta_{10}$ is the ratio of baryon and photon number densities
in units $10^{10}$. A very small baryon-to-photon
ratio makes nucleosynthesis temperature much lower than
typical nuclear binding energies.
A few years ago $\eta$ was determined from
BBN itself and now the measurements of the angular spectrum of CMBR
(see sec. \ref{s-cmbr}) permit to find it independently. According to
the latter $\eta_{10} \approx 5$. This result is in
a reasonable agreement with
determination of $\eta_{10}$ from BBN. 
Once $T_{NS}$ is reached light
elements are very quickly formed and practically all neutrons, that
survived to this moment, were binded in $^4He$. The mass fraction of
this element was about 25\%, while relative number of deuterium nuclei
(as well as $^3He$) with respect to hydrogen is 
(a few)$\times 10^{-5}$. Amount of $^7Li$ is about five orders of
magnitude smaller. Primordial abundances of light elements as 
functions of $\eta_{10}$ are presented in fig. \ref{primabund}.

A slight increase of $^4He$ as a function of $\eta$ 
can be mostly explained
by an increase of the nucleosynthesis temperature (\ref{TNS}) with
rising $\eta$. Correspondingly nucleosynthesis started earlier and more
neutrons survived decay. A strong decrease of the amount of the
produced deuterium is explained by a larger probability for $^2H$ to
meet a nucleon and to proceed to $^4 He$. 

To summarize, we have seen that the primordial abundances depend
upon:
\begin{enumerate}
\item{}
Number density of baryons, $\eta = n_B / n_\gamma$.
\item{}
Weak interaction rate; this is usually expressed in terms of the
neutron life-time. It is interesting that a variation of weak
interaction strength by a factor of few would result either in
complete absence of primordial $^4 He$ or in 100\% dominance of the
latter (no hydrogen). In both cases stellar evolution would be quite
different from what we observe. 
\item{}
Cosmological energy density; non-standard contribution is usually
parameterized as additional number of neutrino species $\Delta N_\nu$. 
\item{}
Neutrino degeneracy; degeneracies of $\num$ or $\nut$ are equivalent to
a non-zero $\Delta N_\nu$, while degeneracy of $\nue$ has a much
stronger (exponential) impact, see eq.~(\ref{n/p}), on the abundances 
because $\nue$ directly enters the (n-p)-transformation reactions
(\ref{nnue},\ref{ne}) 
and can shift the $(n/p)$-ratio in either direction.
\end{enumerate}

\subsection{Role of neutrinos in BBN \label{ss-nubbn}}

The role that neutrinos played in BBN is already clear from the
previous section. The sensitivity of BBN to the number of neutrino
families was first noticed by Hoyle and Tayler in
1964~\cite{hoyle64}. Two years later a similar statement was made by
Peebles~\cite{peebles66}. More detailed calculations were performed by
Shvartsman~\cite{vfs} in 1969 who explicitly stated that the data on 
light element abundances could be used to obtain a bound on the number 
of neutrino flavors. Another 8 years later Steigman, Schramm, and
Gunn presented analysis of the effect with all light elements
taken into account. 

\begin{figure}[htb]
\begin{center}
  \leavevmode
  \hbox{
\epsfysize=4.0in
    \epsffile{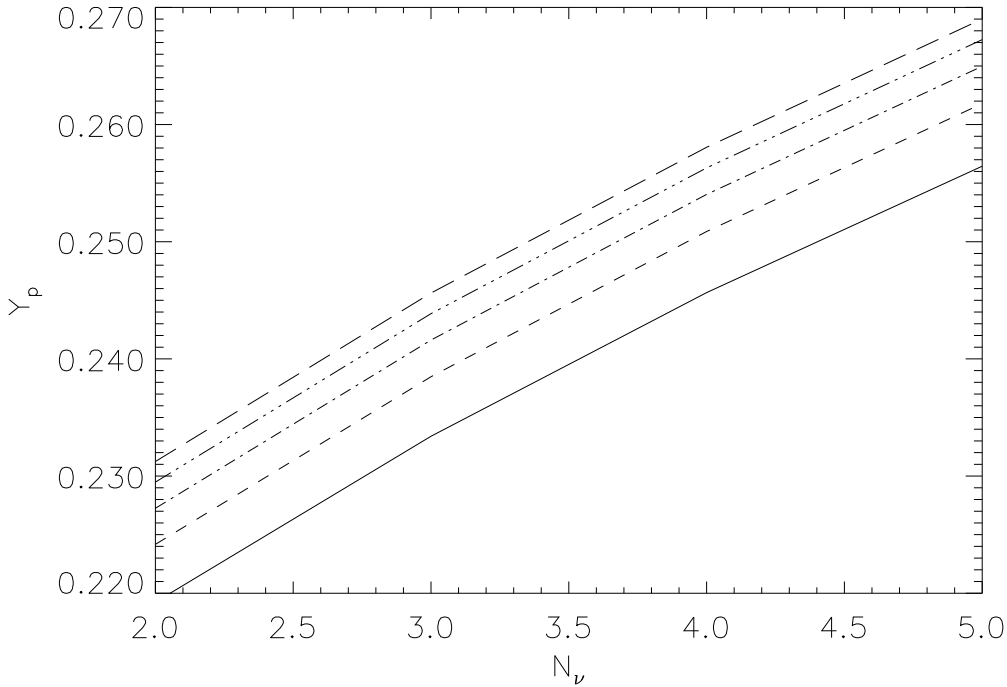}}
\end{center}
\caption{Mass fraction of $^4 He$ as a function of the number of massless
neutrino species. Different curves correspond to different values of the
baryon-to-photon ratio $\eta_{10} \equiv 10^{10}n_B/n_\gamma = 
2,3,4,5,6$ in order of increasing helium abundance.
%2\,{\rm (solid)},\,  3\,{\rm ( dotted)},\,4 \,{\rm (dashed)},\,
%{\rm and}\,\, 5\, {\rm (dashed-dotted)} $.
\label{helknu}}
\end{figure}

\begin{figure}[htb]
\begin{center}
  \leavevmode  
\hbox{
\epsfysize=4.0in    
\epsffile{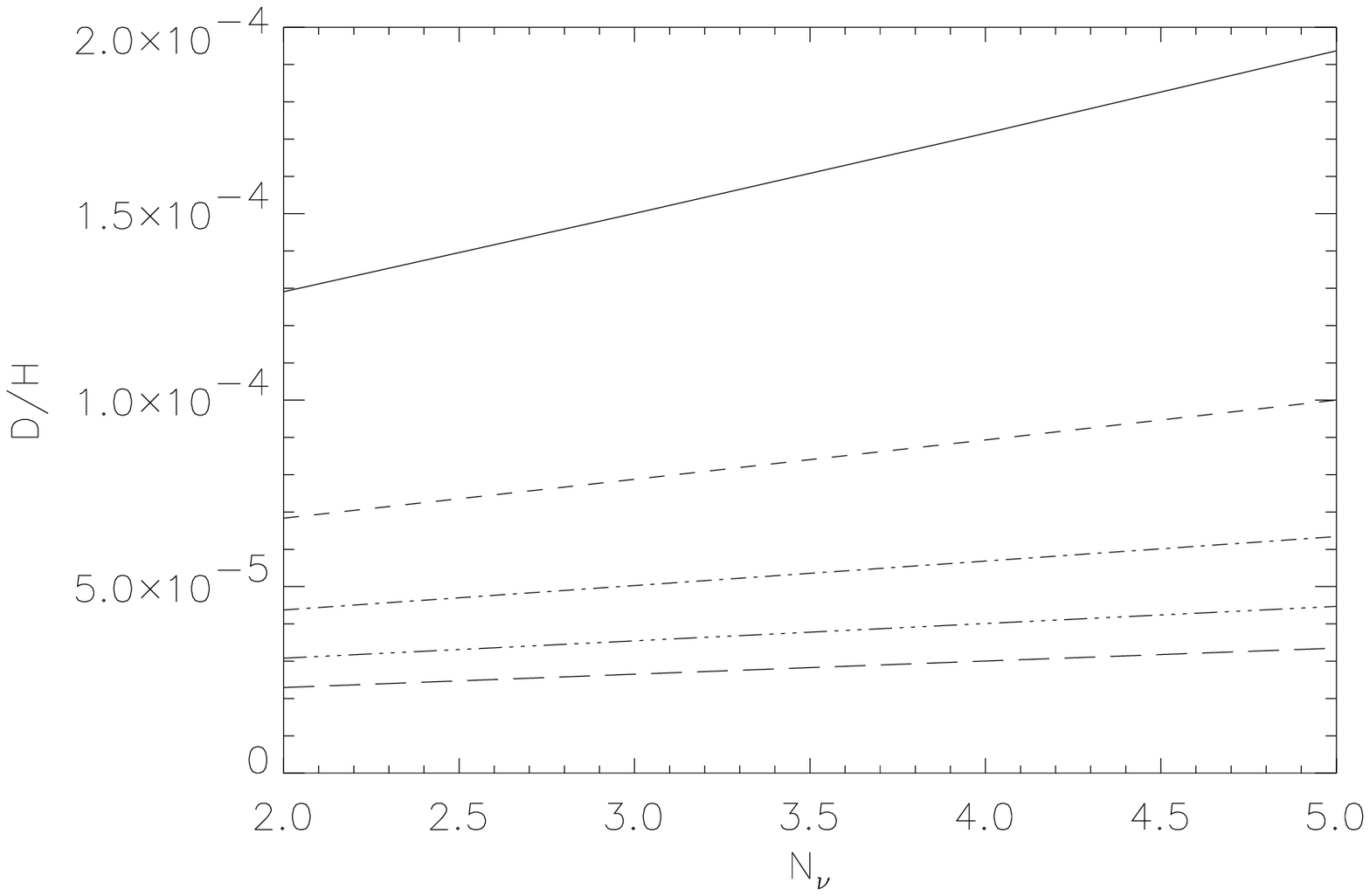}}
\end{center}
\caption{Deuterium-to-hydrogen by number as a function of the number of
massless
neutrino species. Notations are the same as in fig. (\ref{helknu}).
\label{deuterknu}}
\end{figure}

The dependence of the produced deuterium and
helium-4 on the number of neutrino species for different values of the
baryon number density $\eta_{10}$ are presented in 
figs. \ref{helknu},\ref{deuterknu}.
At the present day the conclusion about the allowed number of neutrino
families varies from paper to paper between an optimistic constraint
$\Delta N_\nu < 0.2$ up to a more cautious one $\Delta N_\nu < 1$. The
review of these results can be found e.g. in
refs.~\cite{ad-taup,kainulainen02}. 

If cosmological lepton asymmetry is non-vanishing then chemical
potentials of neutrinos are non-zero and their energy density is
higher than the energy density of
non-degenerate neutrinos. Additional effective number of
neutrino species in this case is given by
\be
\Delta N(\xi) = {15 \over  7}\, \sum_a \left[ 2\,\left(\xi_a \over \pi
\right)^2 + \left(\xi_a \over \pi\right)^2 \right]
\label{deltan-xi}
\ee
where the sum is taken over all neutrino species $a =
e,\,\mu,\,\tau$. If one extra family of neutrinos is allowed by BBN
then chemical potentials of $\num$ and $\nut$ are bounded by
$|\xi_{\mu,\tau}| < 1.5$. For electronic neutrinos the limit is much
stronger, $|\xi_e | < 0.1$~\cite{kohri97}.

If combined variation of all chemical potentials is allowed such that
by some conspiracy an increase in production of light elements due to
non-zero $\xi_{\mu,\tau}$ is compensated by a positive $\xi_e$ then
the limits would be less restrictive. Using additional data from CMBR
which allowed to fix $\eta_{10} = 5$ the authors of 
reference~\cite{hansen01} obtained: 
\be
|\xi_{\mu,\tau}| < 2.6, \,\,\, -0.1 < \xi_e < 0.2
\label{xi-combined}
\ee
This bound disregards mixing between active neutrinos discussed below
in sec. \ref{ss-aa}. Account of the latter may make this limit
considerably more restrictive, eq.~(\ref{xi-aa}).  

Distortion of spectrum of $\num$ or $\nut$ would have an impact on
primordial abundances only through a change in total energy density of
these neutrinos and is equivalent to a change of $N_\nu$. However
distortion of the spectrum of $\nue$ would have a stronger and
non-monotonic influence on the abundances. If there is an increase of
neutrinos in high energy tail of the spectrum
then the frozen $n/p$-ratio would increase. Indeed
an excess of $\nue$ in comparison with the equilibrium amount would
lead to a more efficient destruction of neutrons in reaction
(\ref{nnue}), while the an excess of $\bar\nu$ would lead to more
efficient production of neutrons in reaction (\ref{ne}). If spectral
distortion is symmetric for particles and antiparticles then the second
reaction would dominate because the number density of protons is 6-7
times larger than the number density of neutrons. It is worth noting
that spectral distortion is not necessary charge symmetric as e.g.
could be in the case of resonant oscillations between active and
sterile neutrinos.

If spectrum of $\nue$ has an additional power at low energy then 
the $n/p$ ratio would decrease because the reaction (\ref{ne}), 
where neutrons are
created, is suppressed at low energies due to threshold of 1.8 MeV.

If the spectrum is not distorted but the temperature of neutrinos is
different from the temperature of photons then for a larger 
$T_{\nue}$ the $n/p$-ratio would be smaller because larger
fraction of $\nue$ would shift the neutron freezing
temperature, $T_{np}$, to a smaller value. Thus an increase of the
energy density of $\nue$ has an opposite effect on BBN than an
increase of the energy density of $\num$ or $\nut$.

%\subsection{Heavy unstable neutrinos and BBN \label{ss-heav-unst}}

\section{Neutrino oscillations in the early universe \label{s-nuosc}}

\subsection{Effects of medium \label{medium}}

There are serious reasons to believe that the mass eigenstates of
neutrinos $\nu_j$ ($j=1,2,...$) do not coincide with interaction
eigenstates, $\nu_a$ ($a=e,\mu,\tau,...$). This mismatch leads to
oscillation phenomenon. The interaction and mass eigenstates are
expressed through each other by an unitary mixing matrix which in a
simple case of two particle mixing (e.g. $\nue$ and $\num$) has the 
form:
\be
\nue &=& \nu_1\, \cos \theta + \nu_2\, \sin \theta \nonumber \\
\num &=& -\nu_1\, \sin \theta + \nu_2\, \cos \theta
\label{numix}
\ee
If a certain neutrino flavor (interaction eigenstate) is produced in 
space point $a$ then the probability to observe such state in space
point $b$ oscillates as $\cos [(a-b) \,\delta m^2 /2p]$ where 
$\delta m^2 = m_2^2 -m_1^2$ is the mass difference squared of the mass
eigenstates.    

Oscillations in matter are modified by neutrino effective 
potential $V_{eff}$
which up to energy factor coincides with refraction index of neutrinos
in medium. The Schroedinger equation for neutrino vector wave function
in medium can be written as
\be
i\partial_t \Psi = \left( {\cal H}_m + V_{eff} \right) \Psi
\label{dtpsi}
\ee
where the matrix ${\cal H}_m$ is the free Hamiltonian which is
diagonal in mass eigenstate basis, while effective potential,
$V_{eff}$, is often (but not always, see below, sec. \ref{ss-aa})
diagonal in interaction eigenstate basis. 

Diagonal entries of $V_{eff}$ for an active neutrino were calculated
in ref.~\cite{nora}:
\be 
V_{eff}^{(aa)} = \pm C_1 \eta^{(a)} G_FT^3 + 
C_2^a \frac{G^2_F T^4 E}{\alpha}
\label{veff}
\ee
where $E$ is the neutrino energy, $T$ is the plasma temperature, 
$G_F$ is the Fermi coupling
constant, $\alpha=1/137$ is the fine structure constant, and the signs
``$\pm$'' refer to anti-neutrinos and neutrinos respectively 
According to ref.~\cite{nora} the
coefficients $C_j$ are: $C_1 \approx 0.95$, $C_2^e \approx 0.61$ and
$C_2^{\mu,\tau} \approx 0.17$ (for $T<m_\mu$).  
These values are true in the limit of
thermal equilibrium, otherwise these coefficients are some
integrals from the distribution functions over momenta.
The charge asymmetry of plasma is described by the coefficients
$\eta^{(a)}$ which are equal to
\be
\eta^{(e)} =
2\eta_{\nue} +\eta_{\num} + \eta_{\nut} +\eta_{e}-\eta_{n}/2 \,\,\,
 ( {\rm for} \,\, \nue)~,
\label{etanue} \\
\eta^{(\mu)} =
2\eta_{\num} +\eta_{\nue} + \eta_{\nut} - \eta_{n}/2\,\,\,
({\rm for} \,\, \num)~,
\label{etanumu}
\ee
and $\eta^{(\tau)}$ for $\nut$ is obtained from eq.~(\ref{etanumu}) by
the interchange $\mu \lrar \tau$. The individual charge asymmetries,
$\eta_X$, are defined as the ratio of the difference between
particle-antiparticle number densities to the number density of photons:
\be
\eta_X = \left(N_X -N_{\bar X}\right) /N_\gamma
\label{etax}
\ee

The first term came from thermal averaging of the 
time component of the current with which
neutrino interacts $\langle J_t \rangle$. This operator is odd with
respect to charge conjugation and has different signs for neutrinos
and antineutrinos. The second term appears because of non-locality of
neutrino interaction related to exchange of intermediate bosons.

The numerical values of these two terms in the effective potential as 
well as the energy difference, $\delta E = \delta m^2 /2E$, which 
determines oscillation frequency in vacuum, are 
\be
N^{(a)} &=& C_2^{(a)} G^2_F T^4 E /\alpha = 5.59\cdot 10^{-20}\,
 C_2^{(a)} \left( {T\over {\rm MeV}}\right)^5
\left({ E \over 3T}\right) \,{\rm MeV}
\label{Na} \\
A^{(a)} &=& C_1 \eta^{(a)} G_F T^3 = 1.166\cdot 10^{-21} \,
C_1 \left({\eta^{(a)} \over 10^{-10}}\right)
\left( {T\over {\rm MeV}}\right)^3\,{\rm MeV}
\label{Aa}
\\
\delta E &=& {\delta m^2 \over 2E} = 5\cdot 10^{-13} 
\left( {\delta m^2 \over {\rm eV}^2}\right) \left({{\rm MeV}\over
E}\right) \,{\rm MeV}
\label{deltaE}
\ee  
 
Description of oscillating neutrinos in the early universe in terms of
wave function is not adequate because the effects of breaking of
coherence by neutrino annihilation or non-forward scattering, as well
as neutrino production are essential and one has to use density matrix
formalism~\cite{dolgov81,rho}. Equation for the evolution 
of the density matrix has the form:
\be
{d\rho \over dt} = \left[{\cal H}_1,\rho\right] -
i \left\{ {\cal H}_2,\rho \right\} 
\label{drhodt}
\ee
where ${\cal H}_1$ is the effective Hamiltonian calculated in the
first order in $G_F$ (see eq. (\ref{dtpsi}) and ${\cal H}_2$ is the
imaginary part of the effective Hamiltonian in the second order in
$G_F$. Square brackets mean commutator and curly ones mean
anticommutator. 
The last term turns into the usual collision integral in
non-oscillating case. In many problems this term is approximated as
$-\gamma (\rho - \rho_{eq})$ where $\gamma$ is the effective strength
of interactions and $\rho_{eq}$ is proportional to unit matrix with
the coefficient $f_{eq}$, the latter being the equilibrium
distribution function. Sometimes this approximation reasonably well
describes realistic situation but in many practically interesting
cases this is not so and more accurate form of the coherence breaking
terms should be taken (see e.g. ref.~\cite{dolgov02}).

\subsection{Impact of active-sterile oscillations on BBN
\label{ss-as}} 

All three mentioned above effects of neutrino influence on BBN may be
present in case of mixing of sterile and active neutrinos. 
\begin{enumerate}
\item{}
Production of $\nus$ by oscillations may be efficient enough to
produce noticeable contribution into total cosmological energy density
making $N_\nu > 3$.
\item{}
If MSW resonance transition\cite{msw} is possible a large lepton
asymmetry may be developed in the sector of active neutrinos and if it
happens to be $\nue$-asymmetry it may result in a drastic change in
primordial abundances of light elements.
\item{}
Oscillations may distort energy spectrum of neutrinos because
probability of transformation depends upon their energy. 
\end{enumerate}

We will discuss here mostly the first phenomenon because of continuing
controversy in the literature (see refs.~\cite{dolgov02,kainulainen02}). 
Discussion and references to
effects of spectral distortion and asymmetry generation can be found
in review~\cite{dolgov02}.
 
Let us assume that there is mixing only
between two neutrinos, one active and one sterile, 
and the MSW resonance
condition is not realized (this is so if $\nus$ is heavier than its 
active partner, $\delta m^2 >0$).  
The density matrix is $2\times 2$ 
and has 4 elements which satisfy the following kinetic
equations: 
\be
\dot \rho_{ss} &=&s_2\, \delta E\, I \label{dtss} \\
\dot \rho_{aa} &=& -s_2\,\delta E\, I -\int d\tau |A_{el}|^2 \left[
\rho_{aa} (p_1) f(p_2) - \rho_(p_3) f(p_4) \right] \nonumber \\
&-&\int d\tau |A_{ann}|^2 \left[
\rho_{aa} (p_1)\bar\rho_(p_2)- f(p_3) f(p_4) \right]
\label{dtaa} \\
\dot R &=& W\,I - (1/2) \Gamma\,R \label{dotR} \\
\dot I &=& -W\,R - (1/2) \Gamma\,I  +
(1/2) s_2\, \delta E (\rho_{aa} -\rho_{ss}) \label{dotI}
\ee
where $R$ and $I$ are real and imaginary parts of the non-diagonal
components of the density matrix, $\rho_{as} = R+iI$, 
$\delta E = \delta m^2 /2 E$, 
$s_2 = \sin 2\theta$, $c_2 =\cos 2\theta$, $W= c_2 \delta E +
V_{eff}$ (no-resonance condition means that $W\neq 0$), 
and $\Gamma$ is the total interaction rate of neutrinos with
all other particles. In Boltzmann approximation it is given by the
expression (\ref{hxdf}) with $D = 80 (1+g_L^2+g_R^2)$. Fermi
corrections are calculated in ref.~\cite{dolgov00}. They diminish the
results by 10-15\%.
Integration in eq. (\ref{dtaa}) is taken over the phase space of
particles 2,3, and 4 that participate in the reaction
$ 1+2 \lrar 3+4$.

First we introduce new dimensionless variables, $x$ and $y$, as defined 
after eq. (\ref{kineq}). After that the last two equations 
(\ref{dotR},\ref{dotI}) can be solved analytically as:
\be
\rho_{as} = {i\over 2}\,\int_0^x {dx_1 \over H_1 x_1} s_2\,\delta E\,
\left(\rho_{aa} - \rho_{ss}\right)_1\,
\exp \left[-\int_{x_1}^x {dx_2 \over H_2 x_2} \left( iW + \Gamma/2
\right)_2 \right]
\label{rhoas}
\ee
Here sub-indices 1 and 2 mean that the corresponding functions are
taken at $x_1$ or at $x_2$.

One can estimate that the maximum production rate of sterile neutrinos 
takes place at the temperature~\cite{barbieri90}
\be
T^{\nus}_{prod} = (10-15)\, (3/y)^{1/3}\,
(\dm/{\rm eV}^2)^{1/6}\,\, {\rm MeV}
\label{tprodnus}
\ee
The first number above is for mixing of $\nus$ with $\nue$, 
while the second one is for mixing with $\num$ or $\nut$.
Thus if the neutrino mass difference is not too small 
the production of
sterile neutrinos is efficient when $\Gamma \gg H$. Hence the
integrals in eq.~(\ref{rhoas}) are exponentially dominated by upper
limits and can be easily taken:
\be
\rho_{as} = R+iI = {s_2\,\delta E 
\over 2W - i\Gamma} \, (\rho_{aa} -\rho_{ss})
\label{rhoas-f}
\ee
This is essentially the stationary point approximation which is valid
if $\Gamma$ is sufficiently large. 

Substituting this result into eq.~(\ref{dtss}) we find
\be
Hx\partial_x \rho_{ss} = {\Gamma\over 4}\, \left(\rho_{aa}-\rho_{ss}\right)\,
{  s_2^2 \over \left( c_2^2  + V_{eff} /\delta E \right)^2 +  
\Gamma^2 /4\delta E^2 } 
\label{dxrhoss}
\ee
If we neglect $\rho_{ss}$ in comparison with $\rho_{aa}$ and 
$\Gamma^2 $ in comparison with $W^2$
then we obtain that the
rate of production of sterile neutrinos is 
\be
\Gamma_s = {1\over 4} \,\Gamma_a \sin^2 2\theta_m
\label{gammas}
\ee
where $\theta_m$ is the effective mixing angle in matter. Its
definition is evident from comparison of eqs. (\ref{dxrhoss}) and
(\ref{gammas}). This result is twice smaller than the estimates used
in earlier papers~\cite{barbieri90,kk90,barbieri91} 
and repeated in the recent 
review~\cite{kainulainen02}.
 
Correspondingly the limit on oscillations parameters becomes weaker:
\be
(\dm_{\nue\nus}/{\rm eV}^2) \sin^4 2\theta_{vac}^{\nue\nus} =
3.16\cdot 10^{-5} (g_*(T^{\nus}_{prod})/10.75)^3 (\Delta N_\nu)^2
\label{dmess2}\\
(\dm_{\num\nus}/{\rm eV}^2) \sin^4 2\theta_{vac}^{\num\nus} =
1.74\cdot 10^{-5} (g_*(T^{\nus}_{prod})/10.75)^3 (\Delta N_\nu)^2
\label{dmmuss2}
\ee
If $\Delta N$ is not very small, then a better approximation in the 
bounds above  would be $\ln^2 (1-\Delta N)$
instead of $(\Delta N)^2$.

Possibly these bounds would be even weaker if we take into account
that oscillations conserve total number density of neutrinos and the
latter can be changed only by the neutrino annihilation into
$e^+e^-$-pairs which froze at rather high temperature (see
discussion after eq. (\ref{hxdf}).

In the resonance case a striking effect of a huge rise of lepton
asymmetry of active neutrinos may take place~\cite{ftv} 
leading to asymmetry close to 1. Due to small
fluctuations of baryon asymmetry this huge lepton asymmetry may 
strongly fluctuate at cosmologically large scales~\cite{dibari99b}.
After neutrino decoupling these inhomogeneities would give rise to
significant fluxes of neutrinos. The latter in turn could create
local electric currents with non-zero vorticity which may be sources
of seed magnetic fields~\cite{dolgov-dg}. Turbulent eddies generated
by such flows could also generate gravitational waves potentially
observable in forthcoming LISA mission~\cite{dgn}.

\subsection{Oscillations between active neutrinos \label{ss-aa}}

Normally active neutrinos are in thermal equilibrium even at low
temperatures and oscillations between them do not lead to any change
in their distribution. If however cosmological lepton asymmetry is
non-zero and different for different neutrino flavors then the
oscillations may change it and lead to equality of all
asymmetries. Naively one would expect that in the case of large
asymmetry the mixing angle in matter is very strongly suppressed and
neutrino flavor transformations are absent. This is not the case
however because of a large non-diagonal matrix elements of 
the effective potential. This was first noticed in 
ref.~\cite{pantaleone92} and discussed in detail in series of 
papers~\cite{samuel93}.  A clear description
these phenomena was presented recently in the paper~\cite{pastor01rs}.

The kinetic equations used in the previous section can be easily
modified to apply to this case.
Let us consider for definiteness oscillations
between $\nue$ and $\num$.
One has to take into account the self-interaction processes
$\nue \num \lrar \nue \num$
and $\nue \bar\nue \lrar \num \bar\num$. The refraction index
is determined
by the forward scattering amplitude and since $\nue$ and $\num$ are
considered to be
different states of the same particle one has to include
both processes when there is a $\nue$ with momentum $p_1$ in initial
state and a
$\nue$ or $\num$ with the same momentum in the final state.
The processes of forward
transformation $\nue \lrar \num$ give non-diagonal contributions to
refraction index. Such transformations always exist, even among 
non-oscillating particles, but only in the case of non-vanishing mixing
the non-diagonal terms in the effective potential become observable.

Now effective Hamiltonian has the form:
\be
H_{int}^{(e,\mu)} =
\delta E \left( \begin{array}{cc} h_{ee}  & h_{e \mu} \\
h_{\mu e} & h_{\mu \mu} \end{array} \right) \equiv
{\delta E \over 2}  \left( h_0 + {\bf  \sigma\, h} \right)
\label{hemu}
\ee
where $\delta E = \dm /2E$ and ${\bf \sigma}$ are Pauli matrices.
The elements of the Hamiltonian matrix (\ref{hemu}) are expressed
through the integrals over momenta of the distribution functions of other
leptons in the plasma and, in particular, of the elements of the density
matrix of oscillating neutrinos themselves. 
The structure of
these terms is essentially the same as those discussed above for 
mixing between active and scalar neutrinos, see eq.~(\ref{veff}).
The contribution of self-interaction of neutrinos and antineutrinos also
contains two terms. One originates from non-locality of weak interactions 
and is symmetric with respect to charge conjugation:
\be
{\bf h}_+ = {V_{sym} \over 2\pi^2} \int dy y^3 \left( {\bf P} +\bar{\bf P}
\right).
\label{vsym}
\ee
The second is proportional to the charge asymmetry in the plasma and 
equals 
\be
{\bf h}_- = {V_{asym} \over 2\pi^2} \int dy y^2 \left( {\bf P} -\bar{\bf P}
\right)
\label{vasym}
\ee
 An essential feature,
specific for oscillations between active neutrinos, is the presence
of non-diagonal terms in the Hamiltonian (or in
refraction index). In the
case of large lepton asymmetry in the sector of oscillating neutrinos,
the asymmetric terms in the Hamiltonian strongly dominate and, as a 
result, the suppression 
of mixing angle in the medium, found for $(\nu_a - \nu_s)$-oscillations,
disappears. To be more precise initially the non-diagonal matrix
elements are zero but they quickly rise, if the 
initial asymmetry is not too
high, and soon become very large.

A large contribution of lepton asymmetry into effective potential
permits to solve kinetic equations for density matrix analytically and
find~\cite{dolgov02}:
\be
h'_z = - {1\over 2}\, h_z\, \int d^3 y\,e^{-y}\,{s_2^2 \gamma \over
\langle \gamma\rangle^2 + c^2_2}
\label{h'z}
\ee
where $\gamma = \Gamma/\delta E$ and brackets in the denominator mean
averaging over neutrino spectrum.

The solution of this equation is straightforward. It shows that 
oscillations are not suppressed by matter effects in 
the presence of large lepton asymmetry. A detailed numerical
investigation of oscillations
between three active neutrinos in the early universe in presence of 
a large lepton asymmetry was carried out in the
paper~\cite{dolgov-act}. Similar investigation both analytical and 
numerical was also performed in the papers~\cite{lunardini01}.

\begin{figure}[t]
\centerline{\epsfig{file=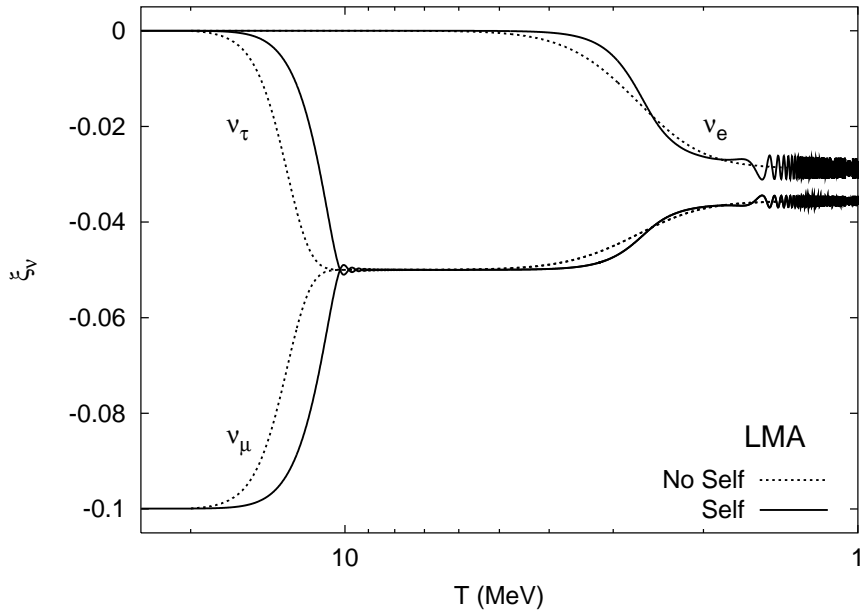,width=12cm}}
\caption{Evolution of neutrino chemical potentials for LMA
case, $\theta_{13}= 0$, and initial values $\xi_e=\xi_\tau=0$ and
$\xi_\mu=-0.1$. Solid and dotted curves are obtained with and
without neutrino self-interactions respectively.
}
\label{f-lma}
\end{figure}

An analysis of the impact of oscillating neutrinos
on BBN was performed for the values of oscillation parameters favored by
the solar and atmospheric neutrino anomalies~\cite{dolgov-act}.
For the large mixing angle
(LMA) solution flavor equilibrium is established in the early universe and
all chemical potentials $\xi_{e,\mu,\tau}$ acquire equal values. The results
of the calculations for this case are presented in fig.~\ref{f-lma}.
Since for these values of
the parameters, asymmetries in muonic and tauonic sectors are efficiently
transformed into electronic asymmetry, the BBN bounds on chemical potentials
are quite strong, 
\be
|\xi_a| < 0.07 
\label{xi-aa}
\ee
for any flavor $a=e,\mu,\tau$.

This result does not destroy the mechanisms of generation of seed magnetic
fields and gravitational waves discussed at the end of sec. \ref{ss-as}.
Indeed, the maximum value of chemical potential that could be
generated by oscillations is 0.6 and may be factor 2-3 smaller 
depending on the mixing parameters. If shared between three neutrino
species these lower values would not contradict the bound
(\ref{xi-aa}). Moreover, for a large region of parameter space the
essential rise of asymmetry takes place below neutron-proton freezing
temperature and does not produce a strong effect on BBN (see
e.g. numerical~\cite{dibari00} or analytical~\cite{dolgov01}
calculations).

\begin{figure}[t]
\centerline{\epsfig{file=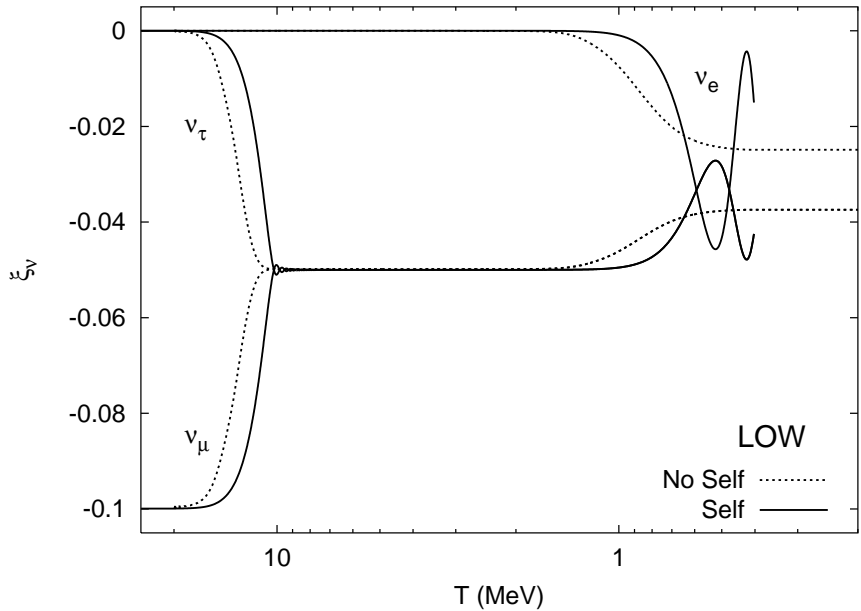,width=12cm}}
\caption{Evolution of the neutrino degeneracy parameters for LOW
case and the initial values $\xi_e=\xi_\tau=0$ and
$\xi_\mu=-0.1$. Notations are the same as in fig.~\ref{f-lma}.
}
\label{f-low}
\end{figure}

For the LOW mixing angle solution the transformation of muon or 
tauon asymmetries into electronic one is not so efficient. The
transformation started at $T<1$ MeV below interesting range for BBN.
The results of calculations are presented in fig.~\ref{f-low}.

The results presented in these figures are valid for 
vanishing mixing angle $\theta_{13}$ (in the standard parameterization of
the $3\times 3$-mixing matrix. An analysis of different non-zero values
of $\theta_{13}$ can be found in the paper~\cite{dolgov-act}.

\bigskip
\bigskip
{\bf Acknowledgment}\\
I am grateful to Yukawa Institute for Theoretical Physics for
hospitality during period when the final version of these lectures was
prepared. I also thank Daniela Kirilova for helpful comments.

\end{document}